\documentclass[aps,prl,reprint,superscriptaddress,showpacs]{revtex4-1}
\usepackage{graphicx}
\usepackage{dcolumn}
\usepackage{bm}
\usepackage[utf8]{inputenc}
\usepackage[T1]{fontenc}
\usepackage{booktabs, array, float, tabularx, booktabs, lipsum, amsmath, multirow}
\usepackage{siunitx, xcolor}
\usepackage[version=4]{mhchem}
\usepackage[colorlinks,linkcolor=blue,anchorcolor=blue,citecolor=blue]{hyperref}
\usepackage{natbib}

\allowdisplaybreaks

\begin{document}

\title{Enhancing strength and range of atom-atom interaction in a coupled-cavity array via parametric drives}
\author{Ya-long Ren}
\affiliation{MOE Key Laboratory for Nonequilibrium Synthesis and Modulation of Condensed Matter and Shaanxi Province Key Laboratory of Quantum Information and Quantum Optoelectronic Devices, School of Physics,
Xi'an Jiaotong University, Xi'an 710049, People's Republic of China}
\affiliation{Physics Division, School of Science and Technology, University of Camerino, I-62032 Camerino, Italy}

\author{Sheng-li Ma}
\email{msl1987@xjtu.edu.cn}
\affiliation{MOE Key Laboratory for Nonequilibrium Synthesis and Modulation of Condensed Matter and Shaanxi Province Key Laboratory of Quantum Information and Quantum Optoelectronic Devices, School of Physics,
Xi'an Jiaotong University, Xi'an 710049, People's Republic of China}

\author{Stefano Zippilli}
\email{stefano.zippilli@unicam.it}
\affiliation{Physics Division, School of Science and Technology, University of Camerino, I-62032 Camerino, Italy}

\author{David Vitali}
\affiliation{Physics Division, School of Science and Technology, University of Camerino, I-62032 Camerino, Italy}
\affiliation{Istituto Nazionale di Fisica Nucleare, Sezione di Perugia, via A. Pascoli, I-06123 Perugia, Italy}
\affiliation{Consiglio Nazionale dell Richerce-Instituto Nazionale di Ottica, L.go Enrico Fermi 6, I-50125 Firenze, Italy}

\author{Fu-li Li}
\affiliation{MOE Key Laboratory for Nonequilibrium Synthesis and Modulation of Condensed Matter and Shaanxi Province Key Laboratory of Quantum Information and Quantum Optoelectronic Devices, School of Physics,
Xi'an Jiaotong University, Xi'an 710049, People's Republic of China}

\begin{abstract}
Coherent long-range interactions between atoms are a prerequisite for numerous applications in the field of quantum information science, but they usually decrease exponentially with the increase in atomic separation. Here we present an appealing method to dramatically enhance the long-range atom-atom interaction mediated by a coupled-cavity array that is subjected to two-photon (parametric) drives. 
Our method allows one to greatly amplify both the localization length of the single-photon bound-state wavefunction and the effective atom-photon coupling strength, resulting in a significant improvement of photon-mediated coherent interaction between two distant atoms. Additionally, we illustrate this effect by analyzing how it facilitates the transfer of information and the creation of entanglement between the atoms.
\end{abstract}

\date{\today}
\maketitle

\emph{Introduction.—}
The pursuit of coherent interactions between atomic emitters, such as neutral atoms, solid-state spins and superconducting qubits,
is a central topic of interest in the quantum information science community \cite{Woerkom2018,Vaidya2018,Evans2018,Borjans2019}.
In fact, it can activate a variety of important applications, including quantum annealing \cite{Glaetzle2017}, quantum sensing \cite{Degen2017}, quantum cryptography \cite{Schimpf2021}, quantum computation \cite{Arute2019} and quantum simulation \cite{Gong2021}.
In particular, strong long-range atom-atom interactions lie at the heart of this subject, unlocking opportunities for long-distance quantum logic gates \cite{Sigillito2019,Marxer2023} and entanglement generation and distribution \cite{Pfaff2014,Newman2018}. Additionally, long-range couplings could trigger a series of exciting phenomena, such as dynamical quantum phase transitions \cite{Zunkovic2018}, non-additivity in statistical mechanics \cite{Campa2009}, exotic long-range order \cite{Maghrebi2017} and the violation of the Lieb-Robinson bound on the speed of information propagation \cite{Richerme2014}. To realize long-range interactions between atoms, a mainstream solution is to interface the atoms with a photonic waveguide and exploit the exchange of virtual photons \cite{Douglas2015,GonzalezTudela2015,Liu2016,Chang2018,Bello2019,Sundaresan2019,Kim2021,Scigliuzzo2022}.
However, in this case, the resulting interactions are exponentially suppressed with the increase of the atomic separation.

In this Letter, we put forward a feasible method to substantially enhance the long-range atom-atom interaction in a coupled-cavity array. We show that, when each photonic site of the array is subjected to a two-photon drive, both the localization length of the single-photon bound-state wavefunction and the effective atom-photon coupling strength can be amplified greatly. Accordingly, when two distant atoms are both coupled to the photonic array in the dispersive regime, the photon-mediated atom-atom interaction is strongly enhanced in terms of both range and strength, pushing it from the weak coupling regime into the strong coupling regime. As a concrete application, we show that, even with a relatively large interatomic distance, quantum entanglement and quantum state transfer between two separated atoms can still be efficiently generated. Finally, we emphasize that our model is quite general and can be implemented with different kinds of architectures, such as superconducting qubits coupled to a microwave cavity array \cite{Liu2016,Sundaresan2019,Kim2021,Scigliuzzo2022,Scigliuzzo2022,Ferreira2021,Zhang2023}
and atomic emitters coupled to an optical cavity array \cite{Thompson2013,Tiecke2014,Reiserer2015,Yu2019,Will2021}.

\begin{figure*}[tbph]
\centerline{\includegraphics[width=14cm]{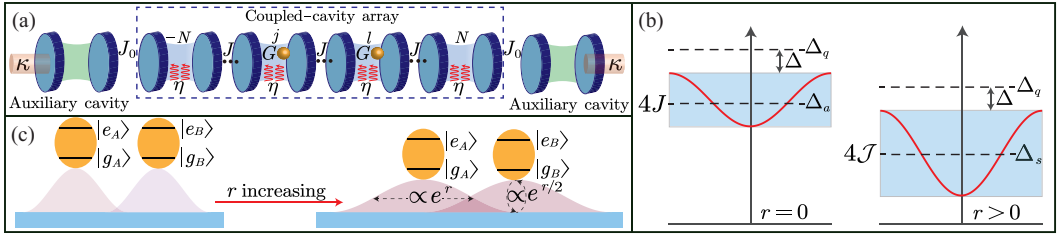}}
\caption{(a) Schematic illustration of the setup: two distant atom are coherently coupled to a coupled-cavity array that is subjected to two-photon drives, and the two edge sites of the array are each coupled to an auxiliary damped cavity.
(b) Propagating frequency band of the coupled-cavity array, which is broadened under the action of two-photon drives.
(c) Schematic diagram of the long-range atom-atom interaction. The parametric drives amplify both
the localization length ($\propto{e^{r}}$) of the single-photon bound-state wavefunction
and the effective atom-photon coupling strength ($\propto{e^{r/2}}$), enabling the long-distance coupling between two atoms.}
\end{figure*}
\emph{The setup.—}
As schematically shown in Fig. 1, we consider a coupled atom-photon system, where two atoms
with excited state $|e_{x}\rangle$ ($x=A$,$B$) and ground state $|g_{x}\rangle$
interact with a one-dimensional coupled-cavity array.
The corresponding Hamiltonian reads (setting $\hbar=1$)
\begin{equation}
\begin{aligned}
H_{a}&=\omega_{a}\sum_{n}a^{\dag}_{n}a_{n}+\omega_{q}\sum_{x}\sigma_{+}^{x}\sigma_{-}^{x} \\
&-\big[J\sum_{n}a^{\dag}_{n}a_{n+1}-G(a^{\dag}_{j}\sigma_{-}^{A}+a^{\dag}_{l}\sigma_{-}^{B})+\mathrm{H.c.}\big],
\end{aligned}
\end{equation}
where $a_{n}$ ($a^{\dag}_{n}$) is the bosonic annihilation (creation) operator of the $n$th cavity with the index $n\in[-N,N]$,
$\sigma_{-}^{x}=|g_{x}\rangle\langle{e_{x}}|$ ($\sigma_{+}^{x}=|e_{x}\rangle\langle{g_{x}}|$)
is the atomic lowering (raising) operator,
and $\omega_{a}$ and $\omega_{q}$ are the resonance frequency of the cavities and atoms, respectively.
$J$ is the cavity-cavity nearest-neighbor coupling strength, and $G$ is the atom-cavity coupling strength.

Moreover, each cavity of the array is subjected to a two-photon drive, described by the Hamiltonian $H_{d}=\eta\sum_{n}(a_{n}^{\dag2}e^{-i\omega_{s}t}e^{-i\phi}+\mathrm{H.c.})$,
where $\eta$ is the two-photon driving amplitude, $\omega_{s}$ is its driving frequency, and $\phi$ is the associated phase.
In the rotating frame at the frequency $\omega_{s}/2$, the Hamiltonian of the system takes the form
\begin{equation}
\begin{aligned}
H&=\Delta_{a}\sum_{n}a^{\dag}_{n}a_{n}+\Delta_{q}\sum_{x}\sigma_{+}^{x}\sigma_{-}^{x}-\big[J\sum_{n}a^{\dag}_{n}a_{n+1} \\
&-\eta\sum_{n}a_{n}^{\dag2}e^{-i\phi}
-G(a^{\dag}_{j}\sigma_{-}^{A}+a^{\dag}_{l}\sigma_{-}^{B})+\mathrm{H.c.}\big],
\end{aligned}
\end{equation}
where $\Delta_{a}=\omega_{a}-\omega_{s}/2$ and $\Delta_{q}=\omega_{q}-\omega_{s}/2$ are the detunings
between the parametric drives and the resonance frequencies of, respectively, the cavities and the atoms.
Here we assume that $\Delta_a$ and $\Delta_q$ are by far the largest parameters in the system dynamics.
This assumption allows us to neglect many non-resonant processes, and the resulting quasi-resonant dynamics shows highly enhanced atom-field interaction similar to the results of \cite{Lue2015,Lemonde2016,Qin2018,Leroux2018}.
Specifically, it is convenient to perform the Bogoliubov squeezing transformations
$a_{n}=\beta_{n}\mathrm{cosh}(r)-\beta_{n}^{\dag}e^{-i\phi}\mathrm{sinh}(r)$
which diagonalize the Hamiltonian of each driven cavity,
where the squeezing parameter is $r=\frac{1}{4}\mathrm{ln}[(\Delta_{a}+2\eta)/(\Delta_{a}-2\eta)]$.

In this squeezed frame, we introduce the modified detuning $\Delta_{s}=\Delta_{a}/\mathrm{cosh}(2r)$, 
and the Hamiltonian (2) can be approximated, by dropping non-resonant terms
in the limit $2\Delta_{s}\gg\mathcal{J}$ and $\Delta_{s}+\Delta_{q}\gg\mathcal{G}$, as
[see details in Supplemental Material]
\begin{equation}
\begin{aligned}
H_{s}&=\Delta_{s}\sum_{n}\beta_{n}^{\dag}\beta_{n}+\Delta_{q}\sum_{x}\sigma_{+}^{x}\sigma_{-}^{x} \\
&-\big[\mathcal{J}\sum_{n}\beta_{n}^{\dag}\beta_{n+1}
-\mathcal{G}(\beta_{j}^{\dag}\sigma_{-}^{A}+\beta_{l}^{\dag}\sigma_{-}^{B})+\mathrm{H.c.}\big],
\end{aligned}
\end{equation}
where $\mathcal{J}=J\mathrm{cosh}(2r)$, and $\mathcal{G}=G\mathrm{cosh}(r)$ are the modified coupling strengths.
Remarkably, the Hamiltonian $H_{s}$ is still described by a tight-binding boson model for the squeezed modes.
Moreover, both the atom-cavity and cavity-cavity couplings can be greatly enhanced
due to the amplified fluctuations of squeezed photons.
This constitutes the key ingredients for increasing the long-distance atom-atom interaction.

So far, we have described a Hamiltonian model without dissipations.
In practice both the cavity array and the atoms inevitably are lossy.
So, when the array is subjected to the two-photon drives,
the driven array approaches a thermal squeezed state in the stationary regime, 
and the associated noise may be detrimental to the atomic dynamics \cite{Qin2018,Leroux2018}.
This problem can be avoided, and the steady state of the cavity array can be approximately prepared into a squeezed vacuum state,
by introducing auxiliary damped cavities at the end of chain [see Fig. 1(a)].
In particular, it is reasonable to assume that the main dissipative channels originate from the auxiliary cavities,
while the dissipation of the internal cavities is comparatively negligible \cite{Scigliuzzo2022}. 
Moreover, we assume that, differently from the cavity array, the auxiliary cavities are not parametrically driven
and they are resonant with the frequency $\Delta_{s}+\omega_{s}/2$ of the squeezed mode
[see details in Supplemental Material]. In this way, they induce dissipation of the corresponding eigenmodes $\beta_{n}$ which are, thereby, driven into their vacuum which, in turn, are pure squeezed states in the original representation.
This means that, with the addition of the auxiliary damped cavities,
it is reasonable to assume that the state of array is the vacuum in the squeezed representation.
In addition, in the dispersive regime, 
the dissipation of the auxiliary cavities has a negligible effect on the atoms for a sufficiently large array.
So, the main dynamics of our system are well described by the Hamiltonian (3) with the array in the squeezed vacuum.
The following analysis exploits this result.

To gain a better understanding of the basic system dynamics,
here we assume that the array is sufficiently large $N\gg1$.
So, we can simplify the model by assuming periodic boundary conditions
for the coupled-cavity array and perform the Fourier transformation
$\beta_{n}=\sum_{k}\beta_{k}e^{-ikn}/\sqrt{2N+1}$ with $k\in[-\pi,\pi]$.
Then, the Hamiltonian $H_{s}$ in Eq. (3) can be transformed into the momentum space
\begin{equation}
\begin{aligned}
H_{s,k}&=\sum_{k}\Delta_{s}^{k}\beta_{k}^{\dag}\beta_{k}+\Delta_{q}\sum_{x}\sigma_{+}^{x}\sigma_{-}^{x} \\
&+\sum_{k}{\mathcal{G}_{k}}(\beta_{k}^{\dag}\sigma_{-}^{A}e^{ikj}+\beta_{k}^{\dag}\sigma_{-}^{B}e^{ikl}+\mathrm{H.c.}),
\end{aligned}
\end{equation}
where $\mathcal{G}_{k}=\mathcal{G}/\sqrt{2N+1}$,
and $\Delta_{s}^{k}=\Delta_{s}-2\mathcal{J}\mathrm{cos}(k)$ is the dispersion relation.

\emph{Enhancing the long-range interaction between two atoms.—}
Let us first consider the coupling of the single atom $A$ to the coupled-cavity array.
We note that, the Hamiltonian (4) conserves the total number of excitations
and as discussed above, it is reasonable to assume that, in the squeezed representation,
the cavities are in their vacuum state, so we can focus on the single-excitation manifold.
Specifically, there exists a bound state
$|\Psi\rangle=\mathrm{cos}(\theta)|\mathrm{Vac}\rangle|e_{A}\rangle+\mathrm{sin}(\theta)\sum_{k}c_{k}|1_{k}\rangle|g_{A}\rangle$
\cite{Douglas2015,Bello2019},
where $\theta$ quantifies the degree of atom-photon hybridization,
$|\mathrm{Vac}\rangle$ is the vacuum state of the modes $\beta_{k}$,
$|1_{k}\rangle=\beta^{\dag}_{k}|\mathrm{Vac}\rangle$ is a single-photon excitation state,
and $c_{k}$ is the amplitude of the single-photon component with momentum $k$.
This single-photon bound state satisfies the eigenvalue equation
$H_{s,k}|\Psi\rangle=\Delta_{BS}|\Psi\rangle$ with the eigenfrequency $\Delta_{BS}$,
which fulfills the relation $\delta=\Delta+\mathcal{G}^{2}/\sqrt{\delta^{2}+4\mathcal{J}\delta}$.
Here, we have introduced the detunings $\delta=\Delta_{BS}-\Delta_{U}$ and $\Delta=\Delta_{q}-\Delta_{U}$
with $\Delta_{U}=\Delta_{s}+2\mathcal{J}$ the upper band edge.
In the following, we focus on the case of $\delta>0$, i.e., $\Delta_{BS}>\Delta_{U}$,
such that the bound state eigenfrequency lies above the propagating frequency band.

The photonic component of the single-photon bound state is localized exponentially around the atomic position.
By utilizing the Fourier transformation $c_{n}=\sum_{n}c_{k}e^{-ikn}/\sqrt{2N+1}$,
we can work out the amplitude $c_n$ of the single-photon component at position $n$
[see details in Supplemental Material]
\begin{equation}
c_{n}=\frac{(-1)^{|n-j|}e^{-\frac{|n-j|}{\xi}}}{\sqrt{\mathrm{coth}(1/\xi)}},
\end{equation}
where the localization length is defined as
\begin{equation}
\xi=\frac{1}{\mathrm{arccosh}(1+\delta/2\mathcal{J})}.
\end{equation}
We can see from Fig. 2(a) that the photonic component extends over multiple sites
and exhibits an exponentially decaying envelope around the atomic position $j$,
which is characterized by the localization length $\xi$.
When we increase the squeezing parameter $r$ at a fixed value of the detuning $\Delta$
between the atomic transition frequency and the upper band edge,
the localization length $\xi$ will be extended exponentially with $\xi\propto{e^{r}}$,
so that the spatial distribution of the localized photon is greatly broadened [see Figs. 2(a) and (b)].

\begin{figure}[tbph]
\centerline {\includegraphics[width=8cm]{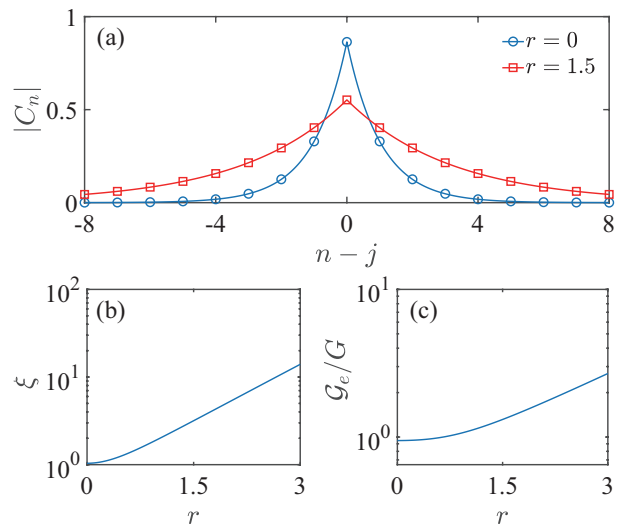}}
\caption{(a) The photonic wavefunction $|c_{n}|$ in the real space versus the spatial position $n-j$.
(b) The localization length $\xi$ and (c) the effective atom–photon coupling strength $\mathcal{G}_{e}$
as a function of the squeezing parameter $r$.
We have set $\Delta=10G$ and $J=10G$.}
\end{figure}

In particular, the photon confined around the single atom $A$
has the same properties as the mode of a real cavity \cite{Chang2018}.
Therefore, the coupling of a single atom to the coupled-cavity array can be well understood
by mapping to the Jaynes-Cummings model with the effective atom–photon coupling strength
\begin{equation}
\mathcal{G}_{e}=\frac{\sqrt{2}\mathcal{G}}{(1+4\mathcal{J}/\delta)^{\frac{1}{4}}},
\end{equation}
and the effective atom-photon detuning $\Delta_{e}=\Delta_{q}-\sum_{k}\Delta_{k}|c_{k}|^{2}=\Delta+\delta/(1+\delta/2\mathcal{J})$.
In Fig. 2(c), we plot the effective atom–photon coupling strength $\mathcal{G}_{e}$ as a function of the squeezing parameter $r$.
As the squeezing parameter $r$ increases,
the effective atom-photon coupling strength will exhibit an exponential enhancement with $\mathcal{G}_{e}\propto{e^{r/2}}$.

Let us now consider two spatially-separated atoms interacting with the coupled-cavity array.
In the dispersive regime $\Delta_{e}\gg\mathcal{G}_{e}$ ($\delta\approx\Delta$),
the effective cavity will mediate a coherent atom-atom interaction via the exchange of virtual photons \cite{Douglas2015}.
By tracing out the modes $\beta_{k}$ in Hamiltonian (4),
we can obtain the effective Hamiltonian in the interaction picture as [see details in Supplemental Material]
\begin{equation}
H_{AB}=\mathcal{G}_{lj}(\sigma_{+}^{A}\sigma_{-}^{B}+\sigma_{-}^{A}\sigma_{+}^{B})
\end{equation}
with
\begin{equation}
\mathcal{G}_{lj}=(-1)^{|l-j|}\frac{\mathcal{G}_{e}^{\prime2}}{2\Delta}e^{-\frac{|l-j|}{\xi^{\prime}}},
\end{equation}
where we have the effective atom–photon coupling strength
$\mathcal{G}_{e}^{\prime}=\sqrt{2}\mathcal{G}/(1+4\mathcal{J}/\Delta)^{\frac{1}{4}}\approx\mathcal{G}_{e}$,
and the localization length $\xi^{\prime}=1/\mathrm{arccosh}(1+\Delta/2\mathcal{J})\approx\xi$.
Note that, the photon-mediated interaction between two atoms naturally inherits the property of the single-photon bound state.
Thus, the resulting atom-atom interaction is exponentially localized
in the real space with $\mathcal{G}_{lj}{\propto}e^{-\frac{|l-j|}{\xi^{\prime}}}$,
\begin{figure}[tbph]
\centerline {\includegraphics[width=8cm]{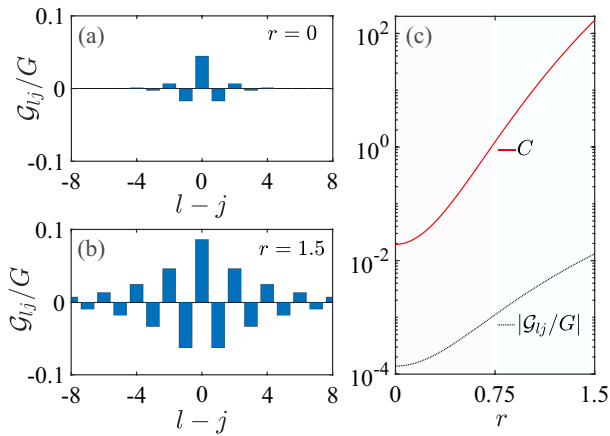}}
\caption{(a,b) The photon-mediated atom-atom coupling strength $\mathcal{G}_{lj}$ as a function of the atomic separation $l-j$.
(c) The absolute value of the coupling strength $|\mathcal{G}_{lj}|$
and the cooperativity $C$ versus the squeezing parameter $r$ for the case of two distant atoms $|l-j|=6$,
where we set $\gamma=0.001G$. The other parameters are the same as in Fig. 2.}
\end{figure}
which is consistent with previous studies
\cite{Douglas2015,GonzalezTudela2015,Liu2016,Chang2018,Bello2019,Sundaresan2019,Kim2021,Scigliuzzo2022}.
However, the unique feature here is that both $\xi^{\prime}$ and $\mathcal{G}_{e}^{\prime}$ increase with the squeezing parameter $r$,
and this will induce a significant enhancement of the photon-mediated atom-atom coupling strength $\mathcal{G}_{lj}$.
To demonstrate this, we plot the coupling strength $\mathcal{G}_{lj}$ as a function of the interatomic distance $l-j$
without and with the parametric driving process in Figs. 3(a) and (b).
For the case of $r=0$, the photon-mediated atom-atom coupling decays rapidly as the atomic separation increases,
which results in a short-distance atom-atom interaction [see Fig. 3(a)].
This is in stark contrast to $r\neq0$,
where the two-photon parametric interactions significantly suppress
the exponential decay behavior of the photon-mediated atom-atom interaction [see Fig. 3(b)].
Hence, even if the two atoms are separated by a relatively large distance,
they can still be strongly coupled to each other.

To quantify the enhancement of $\mathcal{G}_{lj}$,
we further introduce the cooperativity $C=\mathcal{G}_{lj}^{2}/\gamma^{2}$ with the atomic spontaneous emission rate $\gamma$. 
In Fig. 3(c), we plot the photon-mediated atom-atom coupling strength $|\mathcal{G}_{lj}|$
and the cooperativity $C$ as a function of the squeezing parameter $r$ for two distant atoms $|l-j|=6$.
Without the two-photon driving process ($r=0$),
we obtain the relatively small values $|\mathcal{G}_{lj}|\approx1.38\times10^{-4}G$
and $C\approx0.02$, such that the distant atoms are very weakly coupled.
However, these values are strongly enhanced when $r$ is increased, and for $r>0.723$,
we have $|\mathcal{G}_{lj}|>1\times10^{-3}G$ and $C>1$,
i.e., the atom-atom interaction enters into the strong-coupling regime.

\begin{figure}[tbph]
\centerline {\includegraphics[width=8cm]{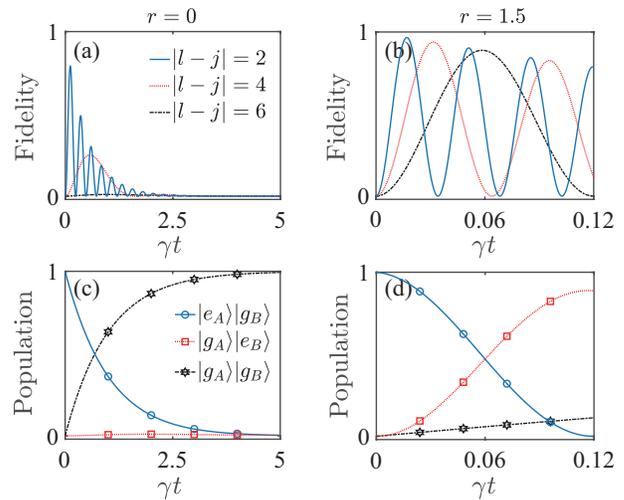}}
\caption{(a,b) Numerical simulations of the fidelity $F$
and (c,d) the populations of two separated atoms versus the time $t$.
The results given in (c) and (d) are for $|l-j|=6$ and the other parameters are the same as in Fig. 3.}
\end{figure}

\emph{Long-range entanglement and quantum information transfer of two atoms.—}
As an application, we now discuss how to achieve long-distance entanglement and quantum information transfer between two separated atoms. If the two atoms initially are prepared in the state $|\phi(0)\rangle=|e_{A}\rangle|g_{B}\rangle$
and the atomic decay is negligible, one can generate a maximally entangled state
$|S\rangle=(|e_{A}\rangle|g_{B}\rangle-i(-1)^{|l-j|}|g_{A}\rangle|e_{B}\rangle)/\sqrt{2}$
at the time $t=\pi/(4|\mathcal{G}_{lj}|)$. Similarly, the state of the atom $A$ in the initial state
$|\phi(0)\rangle=(\alpha_{1}|g_{A}\rangle+\alpha_{2}|e_{A}\rangle)|g_{B}\rangle$,
will be transferred to the atom $B$ with
$|\phi(t)\rangle=|g_{A}\rangle(\alpha_{1}|g_{B}\rangle-i(-1)^{|l-j|}\alpha_{2}|e_{B}\rangle)$
at the time $t=\pi/(2|\mathcal{G}_{lj}|)$.

In the presence of the atomic decoherence, we can use the quantum master equation $\dot{\rho}_{r}=-i[H_{AB},\rho_{r}]+\gamma\sum_{x}\mathcal{L}[\sigma^{x}_{-}]\rho_{r}$
to characterize the dynamics of entanglement and populations of the two separated atoms,
where $\mathcal{L}[\sigma^{x}_{-}]\rho_{r}=[\sigma^{x}_{+}\rho_{r}\sigma^{x}_{-}
-(\sigma^{x}_{+}\sigma^{x}_{-}\rho_{r}+\sigma^{x}_{-}\sigma^{x}_{+}\rho_{r})/2]$
denotes the standard Lindblad operator.
The entanglement of two atoms may be quantified by the fidelity $F=\mathrm{Tr}(\rho_{r}|S\rangle\langle{S}|)$ \cite{2010Quantum}.
Fig. 4 illustrates the numerical results of the time evolution of the fidelity and atomic populations, highlighting the effects of the enhanced cooperativity demonstrated in Fig. 3(c). Specifically, for the chosen parameters, when the atomic distance is large and $r=0$, entanglement generation [see Fig. 4(a)] and population transfer [see Fig. 4(c)] are unfeasible. However, both tasks are made efficient by the two-photon drives, as shown in Figs. 4(b) and 4(d).

\emph{Discussion and conclusions—}
Finally, we discuss the experimental implementations of our scheme.
A promising platform is based on superconducting circuits with their advanced controllability and versatile interfaces \cite{Blais2021}.
Specifically, the cavity array can be made of superconducting microwave cavities embedded with the SQUID
(superconducting quantum interference device) loops,
and the two-photon drive in each cavity can be implemented by modulating the flux threading the SQUID loop \cite{Siddiqi2004,Eddins2019}.
This would allow us to achieve long-range interactions between superconducting qubits
\cite{Liu2016,Sundaresan2019,Kim2021,Scigliuzzo2022,Scigliuzzo2022,Ferreira2021,Zhang2023}
or Rydberg atoms \cite{Morgan2020,Kumar2023GG} coupled to microwave cavities.
Alternatively, in the optical regime, two atomic emitters could be coupled to an array of optical cavities
\cite{Thompson2013,Tiecke2014,Reiserer2015,Yu2019,Will2021}.
In this case, the parametric drives can be realized by degenerate parametric optical down-conversion \cite{Fuerst2010,Fuerst2011,Gao2021}.

In summary, we have proposed an efficient scheme for enhancing long-range interaction between two atoms mediated by a coupled-cavity array. We show that, when each site of the photonic chain is subjected to a parametric drive, the localization length of the single-photon bound-state wavefunction and the effective atom-photon coupling strength are significantly increased, enabling a huge enhancement of the coherent photon-mediated interaction between two distant atoms. So, long-range entanglement and quantum information transfer between two remote atoms can be achieved. Our proposal is general, and can also be applied to other quantum systems, such as phononic \cite{Wang2018,Burd2021} and magnonic crystals \cite{Andrich2017, Kranzl2022}.

\emph{Acknowledgement.—}
We acknowledge financial support from the National Nature Science Foundation of China (Grant Nos. 12074307 and 11704306)
and financial support from NQSTI within PNRR MUR project PE0000023-NQSTI.

\bibliography{RefEn}

\end{document}